\documentclass[12pt,bibliography]{article}
\usepackage{amssymb}
\usepackage{graphicx}
\usepackage{hyperref}
\usepackage{epsfig}
\usepackage{amsfonts}
\usepackage{amsmath}

\newcommand{\ra}{\rightarrow}

\newcommand{\be}{\begin{equation}}
\newcommand{\ee}{\end{equation}}
\newcommand{\ba}{\begin{eqnarray}}
\newcommand{\ea}{\end{eqnarray}}
\newcommand{\bi}{\begin{itemize}}  
\newcommand{\ei}{\end{itemize}}

\newcommand{\Acal}{{\mathcal A}} 
\newcommand{\Bcal}{{\mathcal B}}

\newcommand{\Fcal}{{\mathcal F}}

\newcommand{\Lcal}{{\mathcal L}}

\newcommand{\Ncal}{{\mathcal N}}
\newcommand{\Ocal}{{\mathcal O}}

\newcommand{\Tcal}{{\mathcal T}}

\newcommand{\nn}{\nonumber}

\newcommand{\aslash}[1]{\,\,{\raise.15ex\hbox{/}\mkern-12mu #1}}
\newcommand{\bslash}[1]{\,\,{\raise.15ex\hbox{/}\mkern-9mu #1}}

\renewcommand{\bar}{\overline}

\renewcommand{\Re}{{\rm Re\,}}

\newcommand{\sect}[1]{\section{#1}\setcounter{equation}{0}}

\newcommand\lrpar{\raise .8ex\hbox{$^\leftrightarrow$} \hspace{-9pt}
\partial}
\newcommand\lpar{\raise .8ex\hbox{$^\leftarrow$} \hspace{-9pt}
\partial}
\newcommand\rpar{\raise .8ex\hbox{$^\rightarrow$} \hspace{-9pt}
\partial}

\newcommand{\gsim}{\lower.7ex\hbox{$\;\stackrel{\textstyle>}{\sim}\;$}}
\newcommand{\lsim}{\lower.7ex\hbox{$\;\stackrel{\textstyle<}{\sim}\;$}}

\begin{document}

\baselineskip=18pt

\setcounter{footnote}{0}
\setcounter{figure}{0}
\setcounter{table}{0}

\begin{titlepage}

{\begin{flushright}
 {\bf      NSF-KITP-10-018}
\end{flushright}}

\begin{center}
\vspace{1cm}

{\Large \bf  String scattering in flat space and a scaling limit of Yang-Mills
correlators}

\vspace{0.8cm}

{\bf Takuya Okuda$^1$, Jo\~ao Penedones$^2$}

\vspace{.5cm}

{\it $^1$ Perimeter Institute for Theoretical Physics\\ 
Waterloo, Ontario, N2L 2Y5, Canada}

{\it $^2$ Kavli Institute for Theoretical Physics \\Santa Barbara,
California 93106-4030, USA}

\end{center}
\vspace{1cm}

\begin{abstract}
We use the flat space limit of the AdS/CFT correspondence to derive a simple relation between the $2\to2$ scattering amplitude of massless string states in type IIB superstring theory on ten-dimensional Minkowski space and a scaling limit of the $\Ncal=4$ super Yang-Mills four point functions. We conjecture that this relation holds non-perturbatively and at arbitrarily high energy.
\end{abstract}

\bigskip
\bigskip


\end{titlepage}

\tableofcontents
 
\sect{Introduction and summary} %
Originating as a phenomenological description of hadron scattering,
string theory is defined at the basic level by a prescription
for perturbative scattering amplitudes.
Using the AdS/CFT correspondence \cite{Maldacena:1997re},
one can go beyond a perturbative formulation and define string theory 
in asymptotically AdS backgrounds in terms of the dual gauge theory.
Immediately after the discovery of AdS/CFT,
it was argued \cite{Polchinski:1999ry, Susskind:1998vk} that the flat space limit
of the correspondence should lead to a holographic formulation
of string theory in   Minkowski space.
This idea was explored in \cite{Gary:2009ae} to obtain an explicit relation between scattering
amplitudes of the bulk gravitational theory 
and certain singularities of the CFT correlation
functions at large 't Hooft coupling.
The aim of this paper is to extend this relation
to string scattering amplitudes.

\begin{figure}[]
\begin{center}
\includegraphics[scale=1.4]{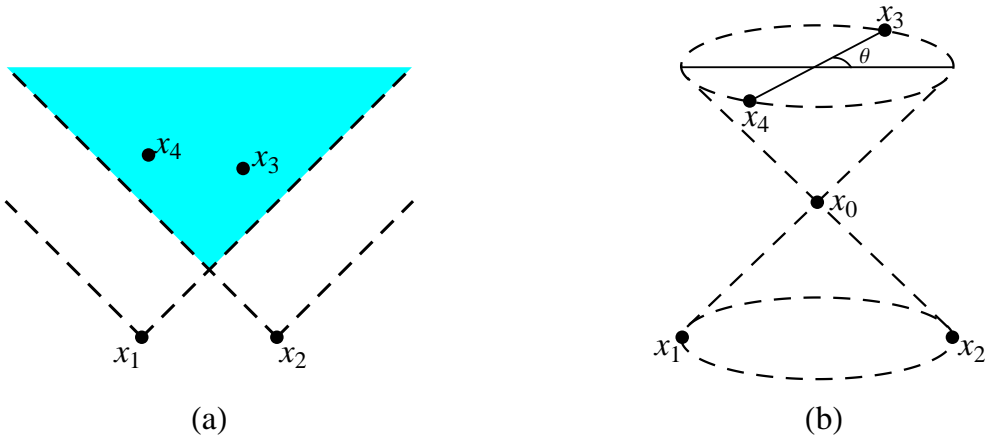}
\end{center}
\caption{
\label{causalrelations}
(a) Causal relations between the operator insertion points used in the flat space limit: $x_{12}$ and $x_{34}$ are spacelike and $x_3$ and $x_4$ are in the future of both $x_1$ and $x_2$.
(b) Example of configuration providing the flat space limit of the correlator. 
The auxiliary point $x_0$ is null related to all the insertion points. 
The conformal invariants for this configuration are $\rho=0$ and $\sigma=\sin^2(\theta/2)$ where $\theta$ is shown in the figure.}
\end{figure}

Relegating the derivation and the more general case to the later sections,
here we  state our main results in the set-up of the  duality between type IIB string theory on $AdS_5\times S^5$
and  $\Ncal=4$ super Yang-Mills (SYM). 
The string coupling $g_s$, the AdS radius $R$ and the string length $l_s$ are related to the Yang-Mills coupling $g_{\rm YM}$ and the number of colors $N$ via
 \be
 4\pi g_s=g_{\rm YM}^2\ ,\ \ \ \ \ \ \ \ \ \ \ \ \ \ \  \left(\frac{R}{l_s}\right)^4 =
  g_{\rm YM}^2N  \ . \nn
 \ee
On the one hand, the natural observables for type IIB string theory on  10D Minkowski space are scattering amplitudes of stable string states. For simplicity, we consider the $2\to 2$ scattering amplitude
of (massless) dilaton particles, $$T=l_s^6\,\Tcal \left(g_s,-t/s ,l_s^2 s\right)\ ,$$
 where $s$ and $t$ are the usual Mandelstam invariants. We recall that the scattering angle $\theta$ is given by $\sin^2 (\theta/2)=-t/s$.
On the other hand, the natural observables for SYM on 4D Minkowski space are correlation functions of gauge-invariant operators. In particular, we are interested in the connected
\footnote{%
The connected four-point function
$\left\langle \Ocal(x_1) \Ocal(x_2)   \Ocal(x_3) \Ocal(x_4) 
\right\rangle_{\rm c}$
is given by subtracting disconnected contributions
from the full four-point function:
$
\left\langle \Ocal(x_1) \Ocal(x_2)   \Ocal(x_3) \Ocal(x_4) 
\right\rangle_{\rm c}
\equiv
\left\langle \Ocal(x_1) \Ocal(x_2)   \Ocal(x_3) \Ocal(x_4) 
\right\rangle
-
\left\langle \Ocal(x_1) \Ocal(x_2)\right\rangle\left\langle  
 \Ocal(x_3) \Ocal(x_4) 
\right\rangle
-\left\langle \Ocal(x_1) \Ocal(x_3) \right\rangle\left\langle 
\Ocal(x_2)   \Ocal(x_4) 
\right\rangle
-\left\langle \Ocal(x_1)\Ocal(x_4)  \right\rangle\left\langle 
 \Ocal(x_2)   \Ocal(x_3) 
 \right\rangle$.
In the text we simply denote $\langle\cdots\rangle_{\rm c}$
by  $\langle\cdots\rangle$.
The two-point functions of BPS operators are not
renormalized \cite{Basu:2004nt}.
} four point function
 of the lagrangian density $\Ocal$, which is the dual operator to the bulk dilaton and is half BPS.
 Conformal invariance implies that
the four-point function on
 Minkowski space $\mathbb M^4$ takes the form
\be
\frac{\left\langle \Ocal(x_1) \Ocal(x_2)   \Ocal(x_3) \Ocal(x_4) 
\right\rangle  }
{\left\langle \Ocal(x_1) \Ocal(x_3) \right\rangle \left\langle \Ocal(x_2) \Ocal(x_4) \right\rangle } =\Acal(g_{\rm YM}^2, N, \sigma, \rho^2) \ ,\nn
\label{reduced-corr}
\ee
where $\sigma$ and $\rho$  are the conformal invariant combinations
\ba 
\sigma^2
=\frac{x_{13}^2x_{24}^2 } {x_{12}^2x_{34}^2}\,, \ \ \ \ 
\sinh^2\rho=
\frac{\det x_{ij}^2}{4\,x_{13}^2x_{24}^2x_{12}^2x_{34}^2} \,. \nn
\ea 
The determinant is taken over $i$ and $j$. We consider $x_{12}=x_1-x_2$ and $x_{34}=x_3-x_4$ spacelike and $x_3$ and $x_4$ in the future of both $x_1$ and $x_2$, as depicted in figure \ref{causalrelations}(a).
We shall see that the flat space limit of the AdS/CFT correspondence
gives
\footnote{%
Our conjectural relations are obtained under the assumption that
the wave packets are localized well enough so that their interactions
are confined to the small flat region in AdS.
A major progress in  \cite{Gary:2009ae} 
was the construction of wave packets 
that are much better localized
than previously considered.
} 
a precise relation between the dimensionless functions $\Tcal$ and $\Acal$.
More precisely, the scaling limit of the four point function,
\be
 \mathcal{F}(g_{\rm YM}^2,\sigma,\xi)=   \lim_{N\to \infty}
\frac{
\Acal\left(g_{\rm YM}^2, N, \sigma, -\frac{ (1-\sigma)\xi^2}{\sigma g_{\rm YM} \sqrt{N} }\right) }
{\sigma^8 \left(g_{\rm YM}^2 N\right)^{7/4}}\ \ \ 
\label{FSlim}
\ee
with
\be
\xi\equiv \left(
-\frac{\sigma}{1-\sigma}\sqrt{g_{\rm YM}^2 N} \rho^2
\right)^{1/2}
\ee
fixed,
is related to the flat space scattering amplitude through the relation
\begin{eqnarray} 
\Fcal(g_{\rm YM}^2,\sigma,\xi)
= \frac{1}{2^{17}3^2\pi^3 \xi \sqrt{\sigma(1-\sigma)}}
\int_0^\infty d\nu \nu^{11} e^{- \xi \nu}
i\Tcal \left(\frac{g_{\rm YM}^2}{4\pi} ,\sigma, \nu^2+i\epsilon\right) 
\ .\label{stringtoCFT}
\end{eqnarray}  
One can also invert this relation and obtain
 \ba
  i \Tcal \left(g_s,-t/s ,l_s^2 s\right)
 =
\frac{2^{17}3^2\pi^3   \sqrt{stu} }{l_s^{11} s^7 }  
  \int^{i\infty}_{-i\infty }
\frac{ d\xi}{2\pi i}  \xi
  \Fcal(4\pi g_s,-t/s,\xi)e^{ \xi l_s \sqrt{s}  }
\,.
\label{CFTtostring}
\ea
Crossing symmetry of the scattering amplitude,
$$
\Tcal  \left(g_s,\sin^2\frac{\theta}{2} ,l_s^2 s\right) 
= \Tcal \left(g_s,\cos^2\frac{\theta}{2} ,l_s^2 s\right)\ ,
$$ 
implies that 
$$\Fcal(g_{\rm YM}^2,\sigma,\xi)=\Fcal(g_{\rm YM}^2,1-\sigma,\xi)\ .
$$
Formulas (\ref{FSlim} - \ref{CFTtostring}),
as well as their generalizations (\ref{xi-general}, \ref{eq:scaling-func}, \ref{FofTgeneral}, 
\ref{TofFgeneral}) to other dimensions,
 are the main results of this paper.
Notice that the dual flat space limit of SYM is a large $N$ limit very different from the planar limit. In the former, one keeps $g_{\rm YM}$ fixed (instead of 
the 't Hooft coupling $g_{\rm YM}^2 N$) and scales the kinematical variable 
$\rho\simeq \det(x_{ij}^2)^{1/2}$ to zero as $ N^{-1/4}$. One way to approach the limit $\rho \to 0$ is to make the four points $x_i$ approach the lightcone of a point $x_0$ in $\mathbb{M}^4$, as in figure \ref{causalrelations}(b). 

The paper is organized as follows.
In Section \ref{sec:derivation},
we derive the relations  (\ref{FSlim} - \ref{CFTtostring})
between the flat space string theory and a scaling limit of gauge theory.
We present two derivations, one of which is in a new approach while the other
is a direct generalization of the methods in \cite{Gary:2009ae}.
Since the relations  (\ref{FSlim} - \ref{CFTtostring})
are rather abstract and involve an unfamiliar limit,
in Section \ref{sec:special-cases} we consider special cases
to gain better intuition.
These limits correspond to the tree-level and hard scattering approximations
on the string theory side.
Finally in Section \ref{sec:discussion}
we discuss the primary obstacles in testing or applying
the relations, and also comment on possible ways to overcome such difficulties.

\sect{Derivations of the relations}
\label{sec:derivation}

In this section we derive
the relations  (\ref{FSlim} - \ref{CFTtostring}).
In fact we present below two derivations, which are
related but independent.
The first one directly
leads to (the generalization of) the relation (\ref{CFTtostring}) that
expresses the scaled CFT correlator
in terms of a string scattering amplitude.
On the other hand the relation (\ref{stringtoCFT})
gives the scattering amplitude as
an integral transform of the CFT correlator.
Since (\ref{stringtoCFT}) mathematically provides the inverse
map of  (\ref{CFTtostring}), logically it suffices to derive only the latter.
In our second derivation, however, we are able to directly obtain
(\ref{stringtoCFT}) by generalizing the analysis performed in
\cite{Gary:2009ae}.

\subsection{From strings to Yang-Mills}

We shall follow the notation of \cite{Gary:2009ae} 
and use the embedding space formalism, which is useful
in relating the analysis in global coordinates in this section
to the expressions
(\ref{reduced-corr}) given in Poincar\'e coordinates of AdS.
A point $X$ in AdS$_{d+1}$ of radius $R$, is described by $X \in \mathbb{R}^{2,d}$ with 
\be
X^2\equiv -(X^{-1})^2-(X^0)^2
+(X^1)^2+\ldots
+(X^d)^2
=-R^2\,.
\ee 
A point in the boundary of AdS is a null ray ($P\sim \lambda P$ and $P^2=0$) in $\mathbb{R}^{2,d}$ \cite{Dirac:1936fq}. 
Correlation functions $$A(P_1,\dots,P_n)= \langle \Ocal_1(P_1)\dots \Ocal_n(P_n)\rangle $$ of scalar primary operators,  
are Lorentz invariant and homogeneous functions on the lightcone of the embedding space $\mathbb{R}^{2,d}$,
$$
A(\dots,\lambda P_i,\dots)=\lambda^{-\Delta_i}A(\dots, P_i,\dots) ,
$$
where $\Delta_i$ is the dimension of the operator at position $P_i$.
Using the AdS/CFT correspondence we can write
\be
A=\int_{\rm AdS} \prod_{i=1}^4 dX_i G_{B\partial}(X_i,P_i) \,G(X_1,\dots,X_4) \label{GreenF}
\ee
where the bulk-boundary propagator is \footnote{
Since supergravity fields like the dilaton are dual to BPS operators whose two and
three point functions are not renormalized, it is very plausible
that their propagators do not receive quantum corrections
 even when the wave length is the string scale.}
\be
G_{B\partial}(X,P)=\frac{C_\Delta}{R^{(d-1)/2}} \frac{1}{(-2P\cdot X /R +i\epsilon)^\Delta} \,,
\ee  
and
$G(X_i)$ is the amputated bulk Green's function
\footnote{%
At intermediate stages of the derivation, we use the Green's function $G$, which is an off-shell field theory quantity hard to define in string theory. However, the final result only depends on the on-shell string scattering amplitude.}.
The normalization constant reads 
$$C_\Delta= \frac{\Gamma(\Delta)}{2\pi^{\frac{d}{2}}
\Gamma\left(\Delta-\frac{d}{2}+1\right)}\ ,$$
and $dX_i$ is the short hand for the volume form on AdS.

Let us now consider a bulk theory with an intrinsic length scale $l_s$ independent of the AdS radius $R$. We are interested in  the limit $l_s \ll R$ so that flat space physics dominates.
In this limit, the important integration region in (\ref{GreenF}) is $|X_i-X_j|  \ll R$ 
\footnote{This condition is clear in the case of massive particles since their propagators decay exponentially with a finite correlation length equal to 
$1/{\rm mass} \sim l_s \ll R$.
For massless particles the situation is more subtle. However, if the diagrams under consideration have no IR divergences in flat space then as we take $R\to \infty$ the integrals become dominated by the flat space region $|X_i-X_j| \ll R$.},
where the AdS curvature effects are negligible. Therefore, we can approximate
\ba
A\approx \int_{\rm AdS} dX_1 G_{B\partial}(X_1,P_1) \int_{\mathbb{M}} \prod_{i=2}^4 dY_i G_{B\partial}(X_1+Y_i,P_i)\,G_{\rm flat}(0,Y_2,Y_3,Y_4) 
\ea
where $Y_i=X_i-X_1$ parametrize the neighborhood
of $X_1$
and $G_{\rm flat}$ is the flat space amputated Green's function. 
Rewriting the bulk-boundary propagator as 
\be
G_{B\partial}(X,P)=\frac{(-i)^{\Delta}C_\Delta R^\Delta}{\Gamma(\Delta) R^{(d-1)/2}
l_s^\Delta}  
\int_0^\infty \frac{d\beta}{\beta} \beta^\Delta e^{-2i\beta P\cdot X /l_s} \ ,\nn
\ee  
we can easily perform the $Y$-integrals and obtain the off-shell flat space scattering amplitude\footnote{To be precise, only the part of $P_i$ orthogonal to $X_1$ contributes. However, this is a subdominant effect.  }
\ba
\int_{\mathbb{M}} \prod_{i=2}^4 dY_i e^{-2i\beta_i P_i\cdot Y_i /l_s}\,G_{\rm flat}(0,Y_2,Y_3,Y_4) 
= iT^{(d+1)}\left(k_1= -\sum_{i=2}^4 k_i, k_i=-2\beta_i P_i /l_s\right)\ . \nn
\ea
The integral over $X_1$ has the form
\ba
\int_{\rm AdS} dX_1 e^{-2i Q \cdot X_1 /l_s} \nn
\ea
with $Q=\sum_{i=1}^4\beta_i P_i$. Using Poincar\'e coordinates, this integral can be simplified to
\ba
-i \pi^{\frac{d}{2}} R^{d+1} \int_0^\infty \frac{dy}{y} (-iy)^{-\frac{d}{2}} e^{iy-
i  (R/l_s)^2 Q^2/y}\nn\ .
\ea
The four point function is then given by
\ba
A &\approx& \pi^{\frac{d}{2}} R^{3-d} \prod_{i=1}^4 \frac{(-i)^{\Delta_i}C_{\Delta_i} R^{\Delta_i}}{\Gamma(\Delta_i)
l_s^{\Delta_i}}   \int_0^\infty \frac{dy}{y} (-iy)^{-\frac{d}{2}} e^{iy}\nn \\
&&\int_0^\infty  \prod_{i=1}^4\frac{d\beta_i}{\beta_i} \beta_i^{\Delta_i} 
e^{-
i  (R/l_s)^2 Q^2/y} \, T^{(d+1)}
\left(k_1= -\sum_{i=2}^4 k_i, k_i=-2\beta_i P_i /l_s\right)
 \ ,\label{betaint}
\ea
with $Q^2=-\frac{1}{2}\sum_{i,j} \beta_i\beta_j P_{ij} $
and $P_{ij}= -2 P_i \cdot P_j$ . 
Since the exponent contains the large factor $ (R/l_s)^2$ we shall attempt to perform the integral over $\vec{\beta}=(\beta_1,\beta_2,\beta_3,\beta_4)$ by saddle point.
This requires diagonalizing  the real symmetric matrix $P_{ij}$. 
In general, this matrix has 4 eigenvalues of order 1 and the integral is small.
However, when one eigenvalue is very small, of order $(l_s/R)^2$, the integral gets enhanced.
This is the kinematical regime of interest which we now describe.
Writing $P=(\cos \tau,\sin\tau, \bf{e})$ we describe the AdS boundary in global coordinates with $\bf{e}$ a $d$-dimensional unit vector parametrizing $S^{d-1}$ and $\tau$ being global time. We then take
\ba
P_1&=&(0,-1,-1,0,\bf{0})+\dots \nn  \\
P_2&=&(0,-1,1,0,\bf{0})+\dots  \label{boundarypoints}\\
P_3&=&(0,1,\cos \theta,\sin \theta,\bf{0})+\dots \nn \\
P_4&=&(0,1,-\cos \theta,-\sin \theta,\bf{0})+\dots  \nn
\ea
where $\bf{0}$ is the origin of $\mathbb{R}^{d-2}$ and the dots stand for small deviations whose explicit form will not be important.
With this choice the matrix $P_{ij}$ has the following eigenvalues and eigenvectors
\be
\begin{array}{lcl}
\lambda_0=0+\dots &\ \ \ \ \ \ \ \ \ \ \   & \vec{\beta}_0=(1,1,1,1)+\dots\\
\lambda_s=8+\dots&  &\vec{\beta}_s=(-1,-1,1,1)+\dots\\
\lambda_t=-8\sigma+\dots& &\vec{\beta}_t=(-1,1,-1,1)+\dots\\
\lambda_u=-8(1-\sigma)+\dots&  &\vec{\beta}_u=(-1,1,1,-1)+\dots
\end{array}\nn
\ee
where $\sigma=\sin^2\frac{\theta}{2}$.
As we shall see, the only important effect of the small deviations from (\ref{boundarypoints}) is to produce a non-vanishing 
eigenvalue $\lambda_0 \neq 0$. In particular, 
$
\det P_{ij} \approx 2^9 \sigma (1-\sigma) \lambda_0.
$

In order to perform the $\vec{\beta}$-integral in (\ref{betaint}) we change coordinates as follows,
\be
4 \vec{\beta}=\nu \vec{\beta}_0 + \nu_s \vec{\beta}_s+
\nu_t \vec{\beta}_t+\nu_u \vec{\beta}_u \ .\nn
\ee
The integrals over $\nu_s,\nu_t$ and $\nu_u$ can be readily performed by saddle point, turning the second line of (\ref{betaint})  into
\ba
\pi^{\frac{3}{2}} 4^{2-\sum \Delta_i} (l_s/R)^3 \frac{(-iy)^{\frac{3}{2}}}{\sqrt{\sigma(1-\sigma)}} \int_0^\infty\frac{d\nu}{\nu} \nu^{\sum\Delta_i -3} 
e^{-i\frac{\xi^2 \nu^2}{4 y}} iT^{(d+1)}(-t/s=\sigma ,l_s^2s=\nu^2)
\ea
where we have introduced the scaling variable
\be
-\xi^2=\lim_{ {\lambda_0 \to 0}\atop {R/l_s \to \infty} } \frac{R^2 \lambda_0  }{2 l_s^2} =
\lim_{ {\det P_{ij} \to 0}\atop {R/l_s \to \infty} }
 \frac{R^2}{l_s^2}\frac{ \det P_{ij}}{
4P_{12}P_{34}\sqrt{P_{13}P_{24}P_{14}P_{23}}} \ . 
\label{xi-general}
\ee
To match the definitions in the Introduction, recall that in the usual Poincar\'e coordinates $P_i=(P^+,P^-,P^\mu)=(1,x_i^2,x_i^\mu)$, one has
$P_{ij}=x_{ij}^2$.

We can now return to (\ref{betaint}) and do the $y$-integral,
\ba
A \approx  \frac{4(2\pi)^{\frac{3+d}{2}}  l_s^3}{R^{d}\sqrt{\sigma(1-\sigma)}} 
\prod_{i=1}^4 \frac{(-i)^{\Delta_i}C_{\Delta_i} R^{\Delta_i}}{\Gamma(\Delta_i)
(4l_s)^{\Delta_i}}   \int_0^\infty\frac{d\nu}{\nu} \nu^{\sum\Delta_i -3} (\xi \nu)^{\frac{3-d}{2}}
K_{\frac{3-d}{2}}(\xi \nu)  iT^{(d+1)}   \label{resultA}
\ea
where $K$ is the modified Bessel function.

Let us now specialize to the case of elastic scattering, where 
$\Delta_1=\Delta_3$ and $\Delta_2=\Delta_4$.
In this case, it is convenient to define a reduced four point function $\Acal$ by dividing $A$ by the disconnected correlator,
$$
A(P_i)=  \frac{C_{\Delta_1}C_{\Delta_2}\,\Acal(P_i) }
{(-2P_1\cdot P_3 +i\epsilon)^{\Delta_1}
(-2P_2\cdot P_4 +i\epsilon)^{\Delta_2}}\ .
$$
The reduced function $\Acal$ here coincides
with $\Acal$ that defined in (\ref{reduced-corr}).
Let us also specialize to the case where the bulk theory is critical string theory on AdS$_{d+1}\times M_{9-d}$. Then, the scattering amplitude $T^{(d+1)}$ is equal to the ten-dimensional string scattering amplitude $T=l_s^6 \Tcal$ divided by the volume $v R^{9-d}$ of the compact space $M_{9-d}$.
Using (\ref{resultA}), we conclude that the scaling function
\be
\Fcal(\xi)=
\lim_{R/l_s \to \infty} \left(\frac{R}{l_s}\right)^{9-2\Delta_1-2\Delta_2}
\frac{\Acal(P_i)}{\sigma^{\Delta_1+\Delta_2}} 
\label{eq:scaling-func}
\ee 
is directly related to the string scattering amplitude via 
\ba
\Fcal(\xi)=\frac{\Ncal}{\sqrt{\sigma(1-\sigma)}}  \xi^{\frac{3-d}{2}}
 \int_0^\infty d\nu\, \nu^{ 2\Delta_1+2\Delta_2 -\frac{5+d}{2}}
K_{\frac{3-d}{2}}(\xi \nu)  i\Tcal (l_s^2s=\nu^2+i\epsilon)\label{FofTgeneral}
\ea
where we have suppressed the $\sigma=-t/s$ and $g_s$ dependence and defined
$$
\Ncal=\frac{(2\pi)^{\frac{3-d}{2}} 2^{d-  2\Delta_1-2\Delta_2} }{v
\Gamma(\Delta_1)\Gamma(\Delta_2)\Gamma\left(\Delta_1-\frac{d}{2}+1\right)
\Gamma\left(\Delta_2-\frac{d}{2}+1\right) }\ .
$$
When $\Delta_1=\Delta_2=d=4$ this Bessel transform reduces to the Laplace transform of equation 
(\ref{stringtoCFT}).

\subsection{From Yang-Mills to strings}

\begin{figure}[]
\begin{center}
\includegraphics[scale=0.6]{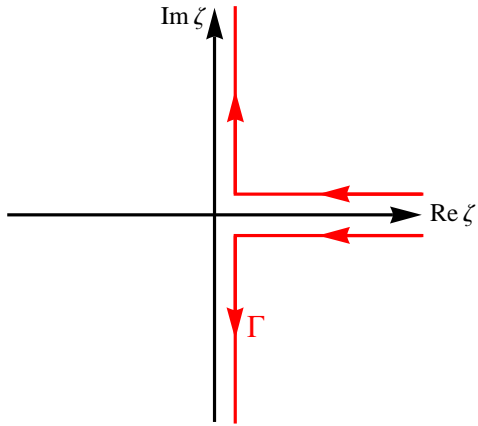}
\end{center}
\caption{\label{BesselContour} Integration contour used in equation \ref{TofFgeneral}.}
\end{figure}

The inverse transform of (\ref{FofTgeneral}), and hence (\ref{CFTtostring})
as a special case,  can be found by using the
wave packet construction of \cite{Polchinski:1999ry,Gary:2009ae}.
Since the set-up is exactly the same as \cite{Gary:2009ae}
and this is the second derivation of the main results,
we will be brief. 
The basic idea was to use sources localized
around the boundary points (\ref{boundarypoints}) 
to produce wave packets  that
scatter in a small flat region of the bulk.
The  operators smeared around $P_1$ and $P_2$
represent incoming particles while
those around $P_3$ and $P_4$
correspond to outgoing particles.
As before, the AdS radius $R$ is taken to be large compared with the string length and the wave length  of the massless particles.
The $(d+1)$-dimensional spacetime momenta of each wave packet is $k_i=(\omega_i,{\bf k}_i)$.
We work in the center-of-mass frame ${\bf k}_1+{\bf k}_2=0$.
Repeating the argument leading to
(3.37)  of \cite{Gary:2009ae},
one can relate the $(d+1)$-dimensional scattering amplitude $T^{(d+1)}$
to $\Acal$ as
\ba
&&
i(2\pi)^{d+1}\delta^{d+1}(\sum k_i)T^{(d+1)}
\nn\\
&&=
\Lcal\,
\delta(\sum\omega_i)
\delta^{2}(\sum {\bf k}_{i||})
\int  d\tau
 e^{-i l_s \omega_1\tau/\sin\theta}
\\
&&
\times
\Acal\left(\sigma, 
\rho^2=\frac{(\tau+i\epsilon)^2/(R/l_s)^2
-
({\bf k}_{4\perp}/\omega_1 )^2
\sin^2\theta}{
16\sin^4(\theta/2)}
\right)
\nn
\ea
up to lower order terms in the limit $R/l_s \ra \infty$.
We have grouped  prefactors into
\ba
\Lcal=\frac{(2\pi)^\frac{d+1}{2}
l_s^{2d-5} (l_s^2 s)^{2-\Delta_1-\Delta_2}
}
{2v \Ncal (\sin(\theta/2))^{2\Delta_1+2\Delta_2}
(R/l_s)^{2\Delta_1+2\Delta_2-2d+2}}.
\nn
\ea
The delta function $\delta^{2}(\sum {\bf k}_{i||})$
enforces momentum conservation in the 
plane spanned by ${\bf k}_1$ and ${\bf k}_3$, and
${\bf k}_{4\perp}$ is the projection of ${\bf k}_4$
perpendicular to that plane.
On the right hand side we 
need an extra delta function $\delta^{d-2}({\bf k}_{4\perp})$,
which arise in the limit $R/l_s\ra \infty$
if $\Acal$ scales as (\ref{eq:scaling-func}).
The coefficient of the delta function can be computed by integrating
over ${\bf k}_{4\perp}$.
We then obtain the relation 
\ba
i\Tcal ( l_s^2s)=\frac{2 l_s^3\sqrt{stu}}{(2 \pi)^{\frac{d+1}{2}} \Ncal}  (l_s^2 s)^{\frac{d-1}{2}-\Delta_1-\Delta_2
}
 \int
d^{d-2}y\int d\tau e^{-i\tau l_s\sqrt{s}}
  \Fcal\left(\xi=\sqrt{y^2-(\tau+i\epsilon)^2}\right) \ ,\nn
\ea
which can be simplified to
\footnote{The consistency of (\ref{TofFgeneral}) and  (\ref{FofTgeneral}) can be checked using 
$$
\int_{\Gamma} d\xi \,\xi K_\alpha(a\xi) K_\alpha(-b\xi) =-\frac{\pi^2}{a}\delta(a-b)\ ,
$$
with $a,b>0$.}
\ba
i\Tcal ( l_s^2s)=-\frac{l_s^3\sqrt{stu}}{\pi^2 \Ncal}  
(l_s^2 s)^{\frac{d+1}{4}-\Delta_1-\Delta_2 }
 \int_{\Gamma} d\xi\,\xi^{ \frac{d-1}{2}}
K_{\frac{d-3}{2}}\left(-\xi l_s\sqrt{s} \right) \Fcal(\xi) \ ,\label{TofFgeneral}
\ea
where $\Gamma$ is the contour shown in the figure \ref{BesselContour}.
Specializing to $d=\Delta_1=\Delta_2=4$  we obtain one
of our main formulas (\ref{CFTtostring}).

\sect{Analysis of special cases}
\label{sec:special-cases}
Now that we have derived the relations
 (\ref{FSlim} - \ref{CFTtostring}) between
the string scattring amplitude and the CFT correlator,
we wish to understand their implications better.
They are highly abstract and given in terms of an unusual scaling limit.
To gain intuition we study two special limits, namely
the tree-level and hard scattering approximations of the string amplitude.

\subsection{Tree-level string theory}%
Let us now focus on the first term in the $g_s$ expansion.
The tree-level  four dilaton scattering amplitude in type IIB string theory is given by \footnote{%
 Recall that 
$ 2\kappa^2=(2\pi)^7 g_s^2 l_s^8
$
and that the polarization tensor for dilaton with momentum $k$ is 
$e_{\mu\nu}= (\eta_{\mu \nu}-k_\mu \bar{k}_\nu -k_\nu \bar{k}_\mu)/\sqrt{8}  $ with $\bar k\cdot k=1, \bar k^2=0$, so that $e_{\mu\nu}e^{\mu\nu}=1$.}
\ba 
  T_{\rm tree}= \kappa^2 \left(
\frac{tu}{s}+\frac{su}{t}+\frac{st}{u}\right) 
\Bcal\left( \frac{ \alpha' s}{4}, \frac{ \alpha' t}{4}, \frac{ \alpha' u}{4}\right) \ \ \ \ 
\ \ \label{treeT}
\ea 
with
$\Bcal(\alpha_1,\alpha_2,\alpha_3)= \prod_{i=1}^3
[\Gamma(1-\alpha_i)/  \Gamma(1+\alpha_i)]
$.
%
Equation (\ref{stringtoCFT}) then gives the planar contribution to the scaling function $\Fcal$
\ba
\Fcal_{\rm planar}= i g_{\rm YM}^4 \frac{\pi^2}{18 \xi} \frac{(1-\sigma+\sigma^2)^2}{\left(\sigma(1-\sigma)\right)^{3/2}}  \int_0^\infty d\nu \nu^{13} \Bcal\left(\nu^2+i\epsilon,-\sigma \nu^2,(\sigma-1)\nu^2\right) e^{-2\xi \nu} \ .\label{Fplanar}
\ea
The small $\alpha'$ expansion of the string scattering amplitude (\ref{treeT}) corresponds to the large $\xi$ expansion, 
\ba
\Fcal_{\rm planar}= i g_{\rm YM}^4 \frac{\pi^2}{9} \frac{(1-\sigma+\sigma^2)^2}{\left(\sigma(1-\sigma)\right)^{3/2}} \left[ 
\frac{\Gamma(14)}{(2\xi)^{15}}+ 
2\zeta(3) \sigma(1-\sigma) \frac{\Gamma(20)}{(2\xi)^{21}} + O\left( \xi^{-25}\right)
\right]\ , \label{largezeta}
\ea
for  $\Re \, \xi >0$. The first term in this expansion can be confirmed by taking the scaling limit of the SYM four point function computed in the supergravity approximation 
\cite{D'Hoker:1999pj}.
The other terms are predictions for the contributions of higher derivative corrections to the spacetime effective action.

The real part of $\Fcal$ is directly related to the imaginary part of the scattering amplitude and therefore signals the presence of production thresholds. 
At tree level, it can be written as
\ba
\Re\, \Fcal_{\rm planar}= - g_{\rm YM}^4 \frac{\pi^3}{36 \xi} \frac{(1-\sigma+\sigma^2)^2}{\left(\sigma(1-\sigma)\right)^{3/2}} \sum_{k=1}^\infty  \frac{(-1)^k k^7}{(k!)^2} 
\frac{\Gamma(1+\sigma k) \Gamma(1+(1-\sigma)k)}
{\Gamma(1-\sigma k) \Gamma(1-(1-\sigma)k)}
  e^{-2\xi \sqrt{k}} \ . \nn
\ea
The exponential decay at large positive $\xi$ means that the real part of $\Fcal$ is undetectable in the small $\alpha'$ expansion (\ref{largezeta}).

\subsection{Hard scattering}  %
It is well-known that string amplitudes are soft at high energy and fixed angle \cite{Gross:1987kza, Gross:1987ar, Mende:1989wt}.
 For instance the amplitude (\ref{treeT}) decays as 
\ba 
  T_{\rm tree}\approx i \kappa^2 s \frac{(1-\sigma+\sigma^2)^2}{\sigma(1-\sigma)}  e^{ -\alpha' s \,q(\sigma)}\ ,
 \label{asymptotics-hard-scattering} 
\ea 
for large $s$ and fixed $\sigma=-t/s$, with 
\ba
-2 q(\sigma)=\sigma\log\sigma+(1-\sigma)\log(1-\sigma)\ . \nn
\ea
Since $ q(\sigma) >0$ for $0<\sigma<1$, 
the integral in (\ref{Fplanar}) converges for all values of $\xi \in \mathbb{C}$, making
$\xi \mathcal{F}_{\rm planar}$ an entire function of $\xi$.
Indeed, we can define 
\ba
\xi \Fcal_{\rm planar}= i g_{\rm YM}^4 \frac{\pi^2}{18 } \frac{(1-\sigma+\sigma^2)^2}{\left(\sigma(1-\sigma)\right)^{3/2}}\sum_{l=0}^\infty a_l(\sigma) \, \xi^l\ ,\nn
\ea
where the series has an infinite radius of convergence.
The high energy asymptotics (\ref{asymptotics-hard-scattering}) gives the large order behavior
\ba 
a_l(\sigma) \approx i  (-1)^l \frac{ \Gamma(7+l/2)}{2^{15} l!\, q(\sigma)^{7+l/2}} \ ,
\ \ \ \ \ \  l\to \infty \ .
\nn 
\ea

At any order in the string perturbative expansion the string amplitude
shows a similar soft behavior in the hard scattering regime. 
More precisely,  the genus $G$ amplitude decays as
   $e^{ -\alpha' s \,q(\sigma)/(1+G)}$ at large $s$.   
This implies that $\xi \mathcal{F}$ is an entire function of $\xi$ at any order in the $g_s$ perturbative expansion. 
The definition (\ref{FSlim}) suggests that this is true even non-perturbatively.
 Furthermore, one expects the contribution of intermediate black hole states to this
$2\to 2$ exclusive process to be suppressed by $e^{-S_{BH}/2}$ \cite{Giddings:2007qq,Giddings:2009gj}.
This fast decay with energy also makes   $\xi \mathcal{F}$ an entire function of $\xi$.

\sect{Discussion} 
\label{sec:discussion}

We derived the conjectural relations
 (\ref{FSlim} - \ref{CFTtostring})
and  their generalizations  (\ref{xi-general}, \ref{eq:scaling-func}, \ref{FofTgeneral}, 
\ref{TofFgeneral})  
using two methods, both on the gravity side of the AdS/CFT duality.
Our relations involve a novel limit of correlation functions,
which is quite natural in the gravity picture and
corresponds to the flat space limit.
Our derivations make use of
the well-localized wave packets
constructed in \cite{Gary:2009ae},
and we believe that the interactions outside the
flat space region do not contribute to the correlation
functions.
%

This limit is however  unusual and hard to analyze on the
field theory side, making it hard to test our relations. 
Obtaining the string scattering amplitude from (\ref{CFTtostring}) requires computing
 the CFT four point function in a Lorentzian and strongly coupled regime.
This is surely a daunting task.
On the other hand, we can use  the relation (\ref{stringtoCFT}) to translate what is 
known about the string amplitude, like
the all-genus calculations of the high-energy 
fixed-angle 
\cite{Gross:1987kza, Gross:1987ar, Mende:1989wt}
and small-angle scattering
\cite{Amati:1987wq, Amati:1987uf},
into predictions
for the gauge theory correlator.

It should also be possible to generalize the relation found here to more general S-matrix elements. In particular, a similar relation should exist between $n$-point functions and $n$-particle scattering amplitudes. We believe this will involve a scaling limit where 
$(R/l_s)^2 \det x_{ij}^2$ is kept fixed while $R/l_s \to \infty$.
Another important issue is the limitation to external massless particles.
Overcoming this would allow us, for instance, to relate the three point function of primary operators to the decay rate of a massive string state into two lighter ones.
 


Clearly the more interesting goal is to find 
a reduction of SYM that computes directly $\Fcal$  or $\Tcal$
without passing through the four point function of the full theory,
in a way rather similar to the BMN limit
\footnote{%
Our limit is also similar to the flat space limit of M(atrix) Theory
\cite{Banks:1996vh} in the sense that both involve strictly infinite $N$.
}.
This would be the long sought holographic dual of flat space.

\section*{Acknowledgements}

We thank A. Buchel, L. Cornalba, M. Costa, J. Gomis, A. Sever, J. Polchinski, R. Porto and 
P. Vieira for useful discussions.
Research at the Perimeter Institute is supported 
in part by the Government of Canada through NSERC 
and by the Province of Ontario 
through MRI.
JP is funded by FCT grant SFRH/BPD/34052/2006. This research was supported in part by the National Science Foundation under Grant No. NSF PHY05-51164. 

\bibliographystyle{utphys}
\bibliography{draft}

\end{document}